\documentclass[12pt,letterpaper]{article}
\usepackage{graphicx}
\usepackage{amsmath}
\usepackage{amssymb}
\usepackage{amsthm}
\usepackage{hyperref}
\usepackage[retainorgcmds]{IEEEtrantools}
\usepackage{collref}

\numberwithin{equation}{section}

\newcommand{\bZ}{\mathbb{Z}}

\newcommand{\bC}{\mathbb{C}}

\newcommand{\CP}{\mathbb{CP}}

\newcommand{\fg}{\mathfrak{g}}

\newcommand{\ft}{\mathfrak{t}}

\DeclareMathOperator{\vol}{vol}
\newcommand{\tgd}{{{\partial}_b{}}}
\newcommand{\tgdb}{{{\bar{\partial}_b}}}

\newcommand{\lineB}{\mathcal{L}}

\DeclareMathOperator{\SU}{SU}
\DeclareMathOperator{\tr}{tr}

\DeclareMathOperator{\id}{id}
\DeclareMathOperator{\ind}{ind}

\DeclareMathOperator{\sdet}{sdet}

\DeclareMathOperator{\sVec}{vec}

\begin{document}

\begin{titlepage}

\vskip 2cm

\begin{center}
{\Large \bfseries
Localisation on Sasaki-Einstein manifolds from holomophic functions on the cone\\
}

\vskip 1.2cm

Johannes Schmude\footnote{schmudejohannes@uniovi.es
}

\bigskip
\bigskip

\begin{tabular}{c}
Department of Physics, Universidad de Oviedo, \\
Avda.~Calvo Sotelo 18, 33007, Oviedo, Spain
\end{tabular}

\vskip 1.5cm

\textbf{Abstract}
\end{center}

\medskip
\noindent
We study super Yang-Mills theories on
five-dimensional Sasaki-Einstein manifolds. Using localisation
techniques, we find that the contribution from the vector multiplet to the
perturbative partition function can be calculated by counting
holomorphic functions on the associated Calabi-Yau cone. This
observation allows us to use standard techniques developed in the
context of quiver gauge theories to obtain explicit results for a
number of examples;
namely $S^5$, $T^{1,1}$, $Y^{7,3}$, $Y^{2,1}$, $Y^{2,0}$, and
$Y^{4,0}$. We find complete agreement with previous results
obtained by Qiu and Zabzine using equivariant indices except for
the orbifold limits $Y^{p,0}$ with $p > 1$.
\bigskip
\vfill
\end{titlepage}

\setcounter{tocdepth}{1}
\tableofcontents

\section{Introduction}
\label{sec:introduction}

Localisation allows for exact
evaluation of path integrals and expectation values of supersymmetric operators \cite{Witten:1988ze,Witten:1991zz,Beasley:2005vf}.
Following the work of Pestun \cite{Pestun:2007rz}, the method has
been applied to a large number of theories in two
\cite{Doroud:2012xw,Benini:2013xpa,Kim:2013ola},
three
\cite{Kapustin:2009kz,Imamura:2011wg,Hama:2011ea,Nian:2013qwa,Tanaka:2013dca,Fujitsuka:2013fga,Alday:2013lba}
four
\cite{Gomis:2011pf,Closset:2013sxa},
and five dimensions, 
\cite{Kim:2012ava,Imamura:2012bm,Kim:2012qf,Kallen:2012cs,Kallen:2012va,Qiu:2013pta,Qiu:2013aga}.
This development went hand-in-hand with an increased interest
in theories with rigid supersymmetry on curved manifolds \cite{Hama:2010av,Hosomichi:2012ek,Festuccia:2011ws,Closset:2012ru,Closset:2013vra}. Eventually, 
K\"allen, Qiu, and Zabzine (KQZ) realised, that the construction of the $S^5$
theory can be directly generalized to generic five-dimensional
Sasaki-Einstein manifolds as it only depends on the existence of the conformal Killing spinors.
Subsequently, the perturbative partition functions of $Y^{p,q}$ and $L^{a,b,c}$
were calculated in \cite{Qiu:2013pta,Qiu:2013aga}. The recent work
\cite{Qiu:2013aga} conjectures the full partition functions using
factorization \cite{Nieri:2013yra,Nieri:2013vba,Lockhart:2012vp}.

The focus of this paper is the perturbative partition function of
vector multiplets on an arbitrary Sasaki-Einstein manifold $Y$. Building on
the work of KQZ we will argue that the one-loop
super determinant can be expressed in terms of the so-called
Kohn-Rossi cohomology groups $H_\tgdb^{p,q}(Y)$ and the Lie-derivative
along the Reeb vector $\xi$.
Previously, the $H^{p,q}_\tgdb(Y)$ have appeared in the context of
holographic calculations of superconformal
indices of three- and four-dimensional SCFTs
\cite{Eager:2012hx,Eager:2013mua,Schmude:2013dua}.
Together with the isomorphism
$H^{0,0}_\tgdb(Y) \cong H^0(\mathcal{O}_{C(Y)})$,
our result allows for a easy evaluation of the perturbative partition function,
as the whole calculation reduces to the counting of
holomorphic functions on the Calabi-Yau cone $C(Y)$, weighted by their
charge along the Reeb. This problem is very well known in the context
of AdS/CFT duality on $AdS \times Y$. Here, the holomorphic functions
on $C(Y)$ correspond to supersymmetric operators in the chiral ring
with $R$-charge determined by $\pounds_\xi$. Following
\cite{Benvenuti:2006qr} we will use the methods developed in this context to evalute
the partition function. To verify our result, we will do explicit
calculations for $S^5, T^{1,1}, Y^{7,3}, Y^{2,1}, Y^{2,0},
Y^{4,0}$. This choice of examples is motivated by the fact that
$Y^{7,3}$ and $Y^{2,1}$ are simple examples of quasi-regular and
irregular Sasaki-Einstein manifolds. We will find full agreement with previous results
except for the last two cases, which arise as $\bZ_2$ and $\bZ_4$ orbifolds of the
conifold. We will comment on this in the conclusions.

While the bulk of this paper uses the methods developed in the context
of AdS/CFT duality to evaluate partition functions, we will reverse 
this logic in the final section. There, we will use the
partition function on $Y^{p,q}$ as well as the insights won from the
examples evaluated to guess the general form of the generating
function for $Y^{p,q}$ written as a series.

The paper is organised as follows. In section \ref{sec:review}, we
review some essentials of Sasaki-Einstein geometry as well as of super
Yang-Mills theories defined on them. Section \ref{sec:main_argument}
contains the main argument of this paper, relating the super
determinant appearing in the perturbative partition function to
Kohn-Rossi cohomology groups. Explicit calcualtion of a number of
examples is done in section \ref{sec:examples} which is complemented
by an evaluation of the same examples using the results of Qiu and
Zabzine for comparison in appendix
\ref{sec:super-determinant_of_Qiu_Zabzine}. The short section
\ref{sec:conjecture_on_generating_functions} concerns the general form
of the generating function for the quiver gauge theories dual to
$AdS_5 \times Y^{p,q}$. Further appendices complement the discussion.

\section{Localisation on Sasaki-Einstein manifolds}
\label{sec:review}

\subsection{Aspects of Sasaki-Einstein geometry}
\label{sec:sasaki-einstein_review}

We begin with a review of the relevant aspects of Sasaki-Einstein
geometry. For a more detailed introduction, we refer to the review
articles \cite{Sparks:2010sn,Boyer:2004fc}; further material on the
tangential Cauchy-Riemann operator and Kohn-Rossi cohomology groups
can be found in \cite{Eager:2012hx,Eager:2013mua,Schmude:2013dua} and
references therein.

Let $Y$ be a five-dimensional Sasaki-Einstein manifold,
$C(Y)$ it's metric cone. $Y$ inherits a number of differential forms
from $C(Y)$; namely the contact form $\eta$, the associated Reeb
vector field\footnote{
In contrast to \cite{Qiu:2013pta,Qiu:2013aga}, we only consider the
Reeb that admits a Sasaki-Einstein metric. We will return to this
restriction in the conclusions.}
$\xi$ satisfying $\xi^\mu \eta_\mu = 1$, a two-form $2 J
= d\eta$, and another two-form $\Omega$. Out of these, only $\Omega$ is
charged under the Reeb:
\begin{equation}
  \pounds_\xi \Omega = 3 \imath \Omega.  
\end{equation}
The tangent bundle $TY$ can be decomposed as $TY = D \oplus L_\xi$,
with $L_\xi$ the line tangent to the Reeb. Moreover, $J$ defines an
endomorphism on $TY$ which satisfies $J^2 = -1 + \xi \otimes \eta$. It
follows that the complexified tangent bundle can be decomposed as
$T_{\bC}Y = (\bC \otimes D)^{1,0} \oplus (\bC \otimes D)^{0,1} \oplus
(\bC \otimes \xi)$. The same holds for the cotangent bundle
\begin{equation}\label{eq:decomposition_of_cotangent_bundle}
  T^*_{\bC}Y = \Omega^{1,0} \oplus \Omega^{0,1} \oplus \bC \eta.
\end{equation}
This decomposition extends to the exterior algebra $\Omega^\bullet =
\bigoplus_{p,q} \Omega^{p,q} \wedge (1 \oplus \eta)$
and to the exterior derivative: $d = \tgd + \tgdb + \eta \wedge
\pounds_\xi$. $\tgdb$ is the tangential Cauchy-Riemann operator. 
Elements of $\bigoplus \Omega^{p,q}$ are sometimes referred to as
horizontal and we will indicate this with a subscript $H$ where
appropriate.
In terms of the decomposition
\eqref{eq:decomposition_of_cotangent_bundle}, the forms $J, \Omega$
are of degree $(1,1)$ and $(2,0)$, while $\eta$ is naturally
transverse to $\Omega^{p,q}$. The complex
\begin{equation}
  \dots \xrightarrow{\tgdb} \Omega^{p,q-1} \xrightarrow{\tgdb}
  \Omega^{p,q} \xrightarrow{\tgdb} \Omega^{p,q+1} \xrightarrow{\tgdb} \dots
\end{equation}
defines the Kohn-Rossi cohomology groups
$H_{\tgdb}^{p,q}(Y)$.

\subsection{The super Yang-Mills theories}
\label{sec:super-yang-mills_review}

We summarize the aspects of \cite{Kallen:2012cs,Kallen:2012va,Qiu:2013pta} that are relevant to
our discussion. Starting point for the construction by Qiu and Zabzine
is the $S^5$ theory of \cite{Hosomichi:2012ek}.
The field content is given by a vector and a hyper multiplet. In this
paper, we will only consider the former. Thus, we are dealing with a
gauge field $A_m$, a scalar $\sigma$, an $\SU(2)$ doublet of scalars
$D_{IJ}$, and a symplectic Majorana gaugino $\lambda_I$. The
Lagrangian is
\begin{IEEEeqnarray}{rCll}
    L_{\sVec} & = & \frac{1}{g_{YM}^2} \tr \bigg( &\frac{1}{2} F_{mn}
    F^{mn} - D_m \sigma D^m \sigma - \frac{1}{2} D_{IJ} D^{IJ} +
    \frac{2}{r} \sigma t^{IJ} D_{IJ} \nonumber\\
    &&& - \frac{10}{r^2} t^{IJ} t_{IJ}
    \sigma^2 + \imath \lambda_I \Gamma^m D_m \lambda^I - \lambda_I
    \lbrack \sigma, \lambda^I \rbrack - \frac{\imath}{r} t^{IJ}
    \lambda_I \lambda_J \bigg).
\end{IEEEeqnarray}
The crucial observation is that both the closure of the
supersymmetry algebra, as well as the invariance of the action depend
only on the existence of the conformal Killing spinors and the dimension
of the space. It follows that the theory as defined on $S^5$ can be
used to define a super Yang-Mills theory on any simply
connected\footnote{This criteria is necessary to ensure the existence
  of the spinors.}
Sasaki-Einstein manifold.

The gaugino is mapped to a $1$-form $\Psi_m$ and a $2$-form
$\chi_{mn}$, the latter of which satisfies
$\imath_\xi \chi = 0$ and $\imath_\xi \star \chi = \chi$.
Using equations \eqref{eq:se_hodge_identities} and
\eqref{eq:horizontal_hodge_identity} one can show that this is
equivalent to the decomposition
\begin{equation}\label{eq:chi_constraints}
  \chi = \chi^{2,0} + \chi^{0,2} + J \chi^{0,0} \quad \text{with} \quad
  \chi^{p,q} \in \Omega^{p,q}.
\end{equation}
Similarly, the $D_{IJ}$ are mapped to a two-form $H$ with an identical
decomposition \eqref{eq:chi_constraints}.
In terms of these, the Localisation term is
\begin{IEEEeqnarray}{rCl}
  V_{\sVec} & = & \tr \left[ \frac{1}{2} \Psi \wedge \star
    (-\imath_\xi F - D \sigma) - \chi \wedge \star H + 2 \chi \wedge
    \star F \right].
\end{IEEEeqnarray}
In the large-$t$ limit the theory localizes to contact instantons,
\begin{equation}
  F_H^+ = 0, \qquad
  \imath_\xi F = 0, \qquad
  D\sigma = 0.
\end{equation}
Here, $F_H^\pm$ is defined as $\frac{1}{2} (1\pm\imath_\xi \star) F_H =
\frac{1}{2} (1\pm\bullet) F_H$, where $\bullet$ is a restriction of
the Hodge dual to horizontal forms and defined in the appendix.

On inclusion of the ghost sector, the perturbative partition function is
\begin{equation}
  \label{eq:Z_pert}
  Z_{\text{pert}} = \frac{1}{\vert W \vert} \frac{\vol (G)}{\vol(T)}
  \int_{\imath \ft} dx \left(\prod_{\beta > 0} \langle \beta, x
    \rangle \right) \exp\left( -\frac{8
      \vol_{\text{SE}}}{g_{\text{YM}}^2 r^2} \tr(x^2) \right)
  \sdet^\prime_{\sVec} (-\imath \pounds_\xi - x)^{1/2}.
\end{equation}
Here, the domain of integration has been reduced from $\fg$ to the
Cartan subalgebra $\ft$ using the Weyl integration formula, $\vert W
\vert$ is the order of the Weyl group, and $\sdet^\prime$ indicates
the exclusion of zero-modes. The modes contributing to the
superdeterminant are as follows:
\begin{equation}
  \begin{array}{ll}
  \text{Bosonic modes}\quad & \Omega^1(Y,\fg) \oplus H^0(Y,\fg) \oplus
  H^0(Y,\fg), \\
  \text{Fermionic modes}\quad & \left(\Omega^{2,0}(Y,\fg) \oplus \Omega^{0,2}(Y,\fg) \oplus
  \Omega^0(Y,\fg)\right) \oplus \Omega^0(Y,\fg) \oplus \Omega^0(Y,\fg).
\end{array}
\end{equation}
The three terms in parantheses are identical to $\Omega^{2+}(Y,\fg)$ in
\cite{Kallen:2012cs}.

\section{One-loop contributions and holomorphic functions}
\label{sec:main_argument}

In this section, we will study the superdeterminant appearing in
\eqref{eq:Z_pert}. To simplify the discussion, we drop the
contribution from the Lie algebra from all expressions. The argument
holds however whether forms are valued in $\bC$ or $\fg$, so one can
simply reinstate them later. We will do so at the end of this section.

Proceeding as in \cite{Kallen:2012cs}, we recall that $\Omega^1 =
\Omega^{1,0} \oplus \Omega^{0,1} \oplus \bC \eta$. The Lie
derivative $\pounds_\xi$ respects this decomposition. As an aside,
note that the tangential Cauchy-Riemann operator $\tgdb$ does
not. Indeed, since $\tgdb \eta = J$, it follows that $\tgdb : \bC \eta
\to \Omega^{1,1} \oplus \Omega^{0,1} \wedge \eta$. In either case,
since the Lie derivative respects the decomposition --- and since the
contact form is not charged under it --- it follows that when it comes
to calculating the determinant, all one-forms can be decomposed into
the sum of a $(1,0)$-, a $(0,1)$-form, and a scalar function. In total,
\begin{equation}
  \label{eq:one-loop_determinant}
  \sdet^\prime(-\imath\pounds_\xi) = 
  \left(
    \frac{\det_{\Omega^0}(-\imath\pounds_\xi)
      \det_{\Omega^{2,0}}(-\imath\pounds_\xi)}{\det_{\Omega^{1,0}}(-\imath\pounds_\xi)}
    \frac{\det_{\Omega^0}(-\imath\pounds_\xi)
      \det_{\Omega^{0,2}}(-\imath\pounds_\xi)}{\det_{\Omega^{0,1}}(-\imath\pounds_\xi)}
  \right)^{\frac{1}{2}}
  \frac{1}{\det_{H^0}(-\imath\pounds_\xi)}.
\end{equation}

In \cite{Kallen:2012cs,Kallen:2012va,Qiu:2013pta}
equation \eqref{eq:one-loop_determinant} is evaluated using index
theorems. We will follow a different route, which is inspired by the
supergravity calculations of \cite{Eager:2012hx,Eager:2013mua}. For
specificity, we focus on the second factor, including determinants
over $\Omega^{0,q}$. For all $f \in \Omega^{0,0}$, we can consider
\begin{IEEEeqnarray}{lr}
  \tgd f \lrcorner \bar{\Omega}, \tgdb f \in \Omega^{0,1}, \qquad &
  f\bar{\Omega} \in \Omega^{0,2}.
\end{IEEEeqnarray}
Any $(0,1)$-form $\alpha$ not included in this list cannot be
$\tgdb$-exact and has to be co-closed. For all such forms, we consider
in addition
\begin{equation}
  \tgdb \alpha \in \Omega^{0,2}.
\end{equation}
One can construct further forms as $\alpha \lrcorner \Omega \in
\Omega^{1,0}$, yet this will be covered by the equivalent discussion
of the factor involving the $\Omega^{p,q}$. Now, schematically,
\begin{equation}
  \frac{\det_{\Omega^0}(-\imath\pounds_\xi)
    \det_{\Omega^{0,2}}(-\imath\pounds_\xi)}{\det_{\Omega^{0,1}}(-\imath\pounds_\xi)}  
  \supset
  (-\imath\pounds_\xi) \left\vert \frac{f}{\tgdb f}
  \frac{f\bar{\Omega}}{\tgd f \lrcorner \bar{\Omega}}
  \frac{\tgdb \alpha}{\alpha}\right. .
\end{equation}
Since $f, \tgdb f$ carry the same charge under $\pounds_\xi$, their
contributions cancel unless $f$ is holomorphic with respect to
$\tgdb$. Identical considerations hold for the other two factors ---
recall that $\pounds_\xi \Omega = 3\imath \Omega$ --- as well as for
the $\Omega^{p,0}$ terms. In the end, we are left with determinants
over Kohn-Rossi cohomology groups
\begin{equation}\label{eq:one-loop-det_as_Kohn-Rossis}
  \sdet^\prime(-\imath\pounds_\xi) = \left( {\det}_{H^{0,0}_\tgdb}^\prime
  (-\imath\pounds_\xi)  {\det}_{H^{0,0}_\tgd}^\prime (-\imath \pounds_\xi)
  {\det}_{H^{0,0}_\tgdb}(-\imath \pounds_\xi - 3)
  {\det}_{H^{0,0}_\tgd}(-\imath \pounds_\xi + 3) \right)^{\frac{1}{2}}.
\end{equation}
The forms $\alpha$ would contribute a determinant over
$H^{0,1}_\tgdb(Y)$, yet this cohomology group vanishes (as does
$H^{1,0}_\tgd(Y)$). Similarly, all harmonic scalar functions on
Sasaki-Einstein manifolds are constants, carry thus zero charge, and
are excluded from the superdeterminant $\sdet^\prime$; so there is no
contribution from $H^0(Y)$ either. The latter follows from the
inequality for the Laplacian, $\Delta \geq -\pounds_\xi^2 - 4 \imath
\pounds_\xi$ proved in \cite{Eager:2012hx,Schmude:2013dua}. Alternatively one can
simply follow the considerations in \cite{Gauntlett:2006vf} in the
context of the discussion of the Lichnerowicz obstruction. There is an
isomorphism $H^{0,0}_\tgdb(Y) \cong H^{2,0}_\tgdb(Y)$, made
explicit by the map $f \mapsto f \Omega$. Since $\Omega$ carries
charge, this absorbs the factors of $3$ in
\eqref{eq:one-loop-det_as_Kohn-Rossis}.
\begin{equation}\label{eq:main_result_using_20-_and_02-forms}
  \sdet^\prime(-\imath\pounds_\xi) = \left( {\det}_{H^{0,0}_\tgdb}^\prime
  (-\imath\pounds_\xi)  {\det}_{H^{0,0}_\tgd}^\prime (-\imath \pounds_\xi)
  {\det}_{H^{2,0}_\tgdb}(-\imath \pounds_\xi)
  {\det}_{H^{0,2}_\tgd}(-\imath \pounds_\xi) \right)^{\frac{1}{2}}.
\end{equation}
In theory, elements of Kohn-Rossi cohomology groups could be
obtained by restriction of corresponding cohomologies on the cone
\cite{Eager:2012hx}. In the case of $H^{0,0}_\tgdb(Y)$ however, we
simply need to count holomorphic functions on the cone:
\begin{equation}
  \label{eq:kohn-rossi_from_cone}
  H^{0,0}_\tgdb(Y) \cong H^0(\mathcal{O}_{C(Y)}).
\end{equation}
Thus, equations \eqref{eq:one-loop-det_as_Kohn-Rossis} and
\eqref{eq:kohn-rossi_from_cone} show that the one-loop contribution to
the partition function \eqref{eq:Z_pert} can be calculated solely in
terms of the holomorphic functions on $C(Y)$.

For the non-abelian case, we follow \cite{Kallen:2012cs,Kallen:2011ny} and decompose
the Lie algebra into root spaces, $\fg = \bigoplus_\beta \fg_\beta$,
which includes the Cartan as $\fg_0 = \ft$. The decomposition extends
to the exterior algebra, $\Omega^{p,q}(Y, \fg) = \bigoplus_\beta
\Omega^{p,q} (Y, \fg_\beta)$. By definition $\forall g
\in \fg_\beta: [ x, g] = \imath\langle\beta, x\rangle g$. Rewriting
$\Omega^{p,q}(Y, \fg_\beta)$ as $\Omega^{p,q}(Y) \otimes \fg_\beta$,
the Lie derivative acts on the first factor while $x$ acts only on the
second. So, in the non-Abelian case we have
\begin{equation}
  \label{eq:non-Abelian_result}
  \sdet^\prime(-\imath\pounds_\xi - x) = \prod_{\beta} \left( {\det_{H^{0,0}_\tgdb}}^\prime
  \left(-\imath O_{\beta, x} \right)  {\det_{H^{0,0}_\tgd}}^\prime \left(-\imath
    O_{\beta, x} \right) \det_{H^{0,0}_\tgdb} \left(-\imath O_{\beta, x} -
    3 \right) \det_{H^{0,0}_\tgd} \left(-\imath O_{\beta, x} + 3\right) \right)^{\frac{1}{2}}
\end{equation}
with $O_{\beta,x} = \pounds_\xi + \langle \beta, x \rangle$.

\section{Examples}
\label{sec:examples}

In this section, we will evaluate
\eqref{eq:one-loop-det_as_Kohn-Rossis} explicitly for a number of
examples and compare our results to those in \cite{Qiu:2013pta}. We
will find complete agreement for $S^5, T^{1,1}, Y^{7,3},
Y^{2,1}$. Curiously, our results disagree for $Y^{2,0}$ and
$Y^{4,0}$. Again we restrict to the Abelian case, keeping in mind that
one can alway incorporate the effect of a non-trivial gauge group as
in equation \eqref{eq:non-Abelian_result}.

Essentially, we will be counting holomorphic functions on $C(Y)$ with
fixed charge under $\pounds_\xi$. Fortunately, this is a very well
understood problem in AdS/CFT duality, due to the following fact:
\emph{Given a Sasaki-Einstein manifold $Y$ and a SCFT dual to
  $AdS\times Y$, the holomorphic functions on $C(Y)$ correspond to
  single trace BPS operators in the chiral ring.}
Note that we are talking about entirely unrelated theories --- four
dimensional SCFTs dual to $AdS\times Y$ with four supercharges and the
five dimensional sYM theories on $Y$ with two supercharges. However,
it should always be entirely clear from the context which theory we
are referring to. Since all our examples are toric, we will be using
\cite{Benvenuti:2006qr,Martelli:2004wu,Martelli:2005tp} to solve the
counting problem. In what follows, we will be looking at generating
functions \cite{Benvenuti:2006qr}
\begin{equation}
  \label{eq:generic_generating_function}
  P(\{t_i\}) = \sum_{i_1, \dots i_k} c_{k_1, \dots, k_n} t^{k_1}_1
  \dots t^{k_n}_n.
\end{equation}
Here, each $t_i$ corresponds to a $U(1)$ symmetry of the SCFT and the
multiplicities $c_{k_1, \dots, k_n}$ count the number of operators
with charge $(k_1, \dots k_n)$. Of course, we are only interested in
the charge under the R-symmetry, so we will set the $t_i$ to the
relevant linear combination as obtained from $a$-maximization
\cite{Martelli:2005tp}.

Given a generating function of the form
\begin{equation}\label{eq:generating_function}
  P(t; C(Y)) = \sum_{n=0} b_n (t^\alpha)^n,
\end{equation}
one finds
\begin{equation}
  \begin{aligned}
    {\det_{H_\tgdb^{0,0}}}^\prime (-\imath \pounds_\xi) &= \prod_{n \geq 1}
    (\alpha n)^{b_n}, &\qquad
    {\det_{H_\tgd^{0,0}}}^\prime (-\imath \pounds_\xi) &= \prod_{n \geq 1}
    (-\alpha n)^{b_n}, \\
    \det_{H_\tgdb^{0,0}} (-\imath \pounds_\xi - 3) &= \prod_{n \geq 0}
    \lbrack \alpha (n - 3/\alpha) \rbrack^{b_n}, &\quad
    \det_{H_\tgd^{0,0}} (-\imath \pounds_\xi + 3) &= \prod_{n \geq 0}
    \lbrack -\alpha (n-3/\alpha)\rbrack^{b_n}.
  \end{aligned}
\end{equation}
The different bounds on the products on the right hand side arise from
the exclusion of zero modes. Thinking in terms of $(0,0)$- and
$(2,0)$-forms, we are excluding constants $c$, yet keeping the form $c \Omega$.
The above can be rewritten by shifting the charge,
\begin{equation}
  \det_{H_\tgdb^{0,0}}(-\imath \pounds_\xi -3) = \prod_{n \geq
    3/\alpha} (\alpha n)^{b_{n-3/\alpha}}, \qquad  
  \det_{H_\tgd^{0,0}}(-\imath \pounds_\xi + 3) = \prod_{n \geq
    3/\alpha} (-\alpha n)^{b_{n-3/\alpha}}.
\end{equation}
When we put everything together, the minus signs will cancel in all
abelian examples and so the overall result for the simpler
cases is
\begin{equation}\label{eq:result_for_simple_cases}
  \sdet^\prime(-\imath \pounds_\xi) = \prod_{n \geq 1} (\alpha n)^{b_n}
  \prod_{n \geq 3/\alpha} (\alpha n)^{b_{n-3/\alpha}}.
\end{equation}

In our examples, we will follow the conventions from
\cite{Martelli:2005tp}. For $Y^{p,q}$, the toric diagram is given by
\begin{equation}\label{eq:toric_diagram_yPQ}
    v_1 = [1,0,0], \quad
    v_2 = [1,p-q-1,p-q], \quad
    v_3 = [1,p,p], \quad
    v_4 = [1,1,0];
\end{equation}
the Reeb vector is
\begin{equation}\label{eq:Reeb_yPQ}
  b_{\min} = \left( 3, \frac{3p-3q+\ell^{-1}}{2},
    \frac{3p-3q+\ell^{-1}}{2} \right);
\end{equation}
and
\begin{equation}\label{eq:ell_inverse}
  \ell^{-1} = \frac{3q^2 - 2p^2 + p\sqrt{4p^2 - 3q^2}}{q}.  
\end{equation}
The generating function of $Y^{p,q}$ is thus
\begin{IEEEeqnarray}{rCl}\label{eq:generating_function_yPQ}
  P(z,x,y;Y^{p,q}) & = \sum_{a=1}^p &
  \frac{1}{(1-yx^{-1})(1-x^{1-a+p-q}y^{a-p+q}z^{1-a})(1-x^{a-p+1}y^{-1-a+p-q}z^a)}
  \nonumber \\
  & &  + \frac{1}{(1-xy^{-1})(1-x^{a-1}y^{2-a}z^{1-a})(1-x^{-a}y^{a-1}z^a)}.
\end{IEEEeqnarray}

\subsection{$S^5$}
\label{sec:examples_s5}

The generating function for $S^5$ can be found in \cite{Benvenuti:2006qr}:
\begin{equation}
  P(t; S^5) = \sum_{n \geq 0} \left( \frac{n^2}{2} + \frac{3n}{2} + 1
  \right) t^n.  
\end{equation}
Upon substitution into \eqref{eq:result_for_simple_cases}, the super
determinant is
\begin{equation}\label{eq:s5_result}
  \sdet^\prime(-\imath \pounds_\xi) = \prod_{n \geq 1} n^{n^2 + 2},
\end{equation}
which agrees with \cite{Kallen:2012cs}.
Since $S^5$ is a $S^1$ bundle over $\CP^2$, one can also use the
Borel-Weil-Bott theorem together with the Weyl dimension formula to
obtain the same result. See appendix \ref{sec:s5-borel-weil-bott}.

\subsection{$T^{1,1}$}
\label{sec:examples_t1-1}

We proceed by considering the next canonical example -- the
base of the conifold ($T^{1,1} = Y^{1,0}$). Here, the generating
function is
\begin{equation}
  P(t; T^{1,1}) = \sum_{n\geq 0} (n+1)^2 \left(t^{3/2}\right)^n.
\end{equation}
Therefore
\begin{equation}\label{eq:t11_result}
  \sdet^\prime(-\imath\pounds_\xi) = \prod_{n \geq 1} \left(\frac{3}{2} n\right)^{2(n^2+1)}.  
\end{equation}
As shown in appendix \ref{sec:super-determinant_of_Qiu_Zabzine}, this
agrees with \cite{Qiu:2013pta}. Both $S^5$ and $T^{1,1}$ are
regular Sasaki-Einstein manifolds.

\subsection{$Y^{7,3}$}
\label{sec:examples_y7-3}

In contrast to the previous two examples, $Y^{7,3}$ is a quasi-regular
Sasaki-Einstein manifold. The condition for quasi-regularity is that
$4 p^2 - 3q^2 = n^2$ with $n \in \bZ$. $(7,3)$ is the simplest
example, followed by $\{ (7,5); (13,7); (13;8); (14,6); (14,10); \dots\}$.
Using \eqref{eq:Reeb_yPQ} and \eqref{eq:generating_function_yPQ}, we
substitute $z\mapsto t^3$, $x, y \mapsto t^{\frac{28}{3}}$. Since
\eqref{eq:generating_function_yPQ} contains terms of order $(1-x/y)^{-1}$ and
our substitution sets $x = y$, one has to take some care when taking
the limit. After doing so, one obtains a series expansion with integer
coefficients in terms of $\tau = t^{\frac{1}{3}}$:
\begin{equation}
  P(\tau; Y^{7,3}) = 1 + 3 \tau^9 + 5 \tau^{18} + 7 \tau^{27} + 5
  \tau^{28} + 11 \tau^{35} + 9 \tau^{36} + 7 \tau^{37} + \mathcal{O}(\tau^{44}).
\end{equation}
There are different ways of rewriting this in the form of equation
\eqref{eq:generating_function}. In order to be able to compare our
result with appendix \ref{sec:super-determinant_of_Qiu_Zabzine}, we define
\begin{equation}\label{eq:y73_generating_function}
  \begin{aligned}
    I_7 &= \{(i, j) \in \bZ^2_{\geq 0} \vert i-j = 0 \mod 7\}, \\
    m_{ij} &= \frac{10i+4j}{7} + 1, \\
    P_\Sigma(\tau; Y^{7,3}) &= \sum_{I_7} m_{ij} \tau^{5i+4j}.
  \end{aligned}
\end{equation}
Using Mathematica, one sees that $P(\tau; Y^{7,3}) - P_\Sigma(\tau;
Y^{7,3}) = \mathcal{O}(\tau^{4001})$ which seems sufficient to assume
that equality holds to all orders and that both series have the same
limit. After some further algebra one finds
\begin{equation}\label{eq:y73_result}
  \sdet^\prime(-\imath\pounds_\xi) = \prod_{I_7 \vert i, j > 0} \left(
    \frac{5i+4j}{3} \right)^{2 \frac{10i+4j}{7}}
  \prod_{I_7 \vert i = 0 \lor j = 0} \left( \frac{5i+4j}{3}
  \right)^{\frac{10i+4j}{7} +1}.
\end{equation}
Again, this agrees with \cite{Qiu:2013pta}.

\subsection{$Y^{2,1}$}
\label{sec:examples_y2-1}

Finally, we turn to an example of an irregular Sasaki-Einstein
manifold, $Y^{2,1}$. The necessary steps are in principle the same as
for $Y^{7,3}$, yet the series expansion is naively a bit more
difficult due to the appearance of irrational exponents. We proceed
by calculating $P(z,x,y;Y^{2,1})$, substituting $y \mapsto x$ and
then performing a double series expansion in $x, z$. In detail,
\begin{equation}
  \begin{aligned}
    P(z,x,x;Y^{2,1}) = \frac{x \{ x - z [ -2z + (-3 +z (3+z))x + 2x^2]
      \}}{(z^2 - x)^2 (1-x)^2}.
  \end{aligned}
\end{equation}
In analogy to section \ref{sec:examples_y7-3}, we define
\begin{equation}\label{eq:y21_generating_function}
  \begin{aligned}
    I_2 &= \{(i, j) \in \bZ^2_{\geq 0} \vert i-j = 0 \mod 2\}, \\
    m_{ij} &= \frac{3i+j}{2} + 1, \\
    P_\Sigma(z,x,x; Y^{2,1}) &= \sum_{I_2} m_{ij} x^{\frac{j-i}{2}} z^i.
  \end{aligned}
\end{equation}
Again, one can check agreement between $P_\Sigma(z,x,x;Y^{2,1})$ and
$P(z,x,x;Y^{2,1})$ using Mathematica; one finds
$P(z,x,x;Y^{2,1}) - P_\Sigma(z,x,x;Y^{2,1}) = \mathcal{O}(z^{3\times150}) \mathcal{O}(x^{(\sqrt{13}-1) \times 150})$.
Now, we
substitute $z \mapsto t^3, x\mapsto t^{\sqrt{13}-1}$ using
\eqref{eq:Reeb_yPQ} and find the generating function of $Y^{2,1}$ in
terms of the Reeb
\begin{equation}
  P_\Sigma(t; Y^{2,1}) = \sum_{I_2} m_{ij} t^{\frac{(7-\sqrt{13})i + (\sqrt{13}-1)j}{2}}.
\end{equation}
Again, this allows us to calculate the one-loop contribution to the
partition function,
\begin{IEEEeqnarray}{rCl}\label{eq:y21_result}
  \sdet^\prime(-\imath\pounds_\xi) = & \prod_{I_2 \vert i, j > 0} & \left(
    \frac{(7-\sqrt{13})i + (\sqrt{13}-1)j}{2} \right)^{3i+j} \nonumber\\
  &\prod_{I_2 \vert i = 0 \lor j = 0}& \left( \frac{(7-\sqrt{13})i + (\sqrt{13}-1)j}{2}
  \right)^{\frac{3i+j}{2} +1}.
\end{IEEEeqnarray}
Once again, appendix \ref{sec:super-determinant_of_Qiu_Zabzine} shows
that this agrees with \cite{Qiu:2013pta}. For a detailed discussion of
$Y^{2,1}$ in the context of quiver gauge theories see \cite{Bertolini:2004xf}.

\subsection{$Y^{p,0}$}
\label{sec:examples_yp-0}

Recall that $Y^{p,0} = (\text{conifold})/\bZ_p$ while
$Y^{p,p} = (\bC^2/\bZ_2 \times \bC) / \bZ_p$
\cite{Martelli:2004wu}. In regards to what follows, one should keep in
mind that it is not clear whether the super Yang-Mills theory is well defined on
orbifolds. Nevertheless, one can use identical methods as in the previous
paragraphs to evaluate the super determinant for $Y^{2,0}$. One finds
\begin{equation}
  \label{eq:y20_result}
  \begin{aligned}
    P(t; Y^{2,0}) &= \sum_{n=0}^\infty (2n+1)^2 (t^3)^2, \\
    \sdet^\prime(-\imath \pounds_\xi) &= \prod_{n \geq 1} (3n)^{2 [(2n)^2 + 1]}
    = \prod_{n \in 2\bZ_{>0}} \left( \frac{3}{2} n \right)^{2(n^2+1)}.
  \end{aligned}
\end{equation}
As to $Y^{4,0}$,
\begin{equation}
  \begin{aligned}
    P(t; Y^{4,0}) &= \sum_{n=0}^\infty
    \frac{(2n+1)(2n+1 - (-1)^{n+1})}{2} (t^3)^n, \\
    \sdet^\prime(-\imath \pounds_\xi) &= \prod_{n \geq 1} 
    (3n)^{4n^2 + (-1)^n + 1}.
  \end{aligned}
\end{equation}
As we argue in appendix
\ref{sec:super-determinant_of_Qiu_Zabzine}, if one naively applies the
results of \cite{Qiu:2013pta} for the one-loop contribution on
$Y^{p,0}$, the result is always \eqref{eq:t11_result}, independent of
$p$. Clearly, both our results for $Y^{2,0}$ and $Y^{4,0}$ do not show
this behavior. While in the former case the result has the same
overall form with the product being taken over a different lattice,
this is not the case for $Y^{4,0}$. Since the result for $Y^{2,0}$
differs from the $Y^{1,0}$ one by a factor two in the lattice spacing,
one can speculate whether the two will agree after
renormalization. Naive application of zeta function regularization
does not yield agreement.

\section{Generating functions for $Y^{p,q}$}
\label{sec:conjecture_on_generating_functions}

So far, we have used \cite{Benvenuti:2006qr} and
\cite{Martelli:2005tp} in order to compute
\eqref{eq:one-loop-det_as_Kohn-Rossis} and compare the result with
that of \cite{Qiu:2013pta}. In this section, we simply invert this
process and use the general form of the contribution to the one-loop partition
function from \cite{Qiu:2013pta} in order to guess the generating
function for generic $Y^{p,q}$ manifolds in terms of the Reeb; i.e.~as
in equation \eqref{eq:generating_function}. While
\cite{Benvenuti:2006qr} gives a prescription for the calculation 
of generating functions that is very straightforward to implement,
rewriting them in the form \eqref{eq:generating_function} can be a bit of a nuisance, as our
calcualtions for $Y^{7,3}$ and $Y^{2,1}$ show. Thus, comparing our
results \eqref{eq:y73_generating_function} and
\eqref{eq:y21_generating_function} with the material in appendix
\ref{sec:super-determinant_of_Qiu_Zabzine} suggests that
\begin{equation}\label{eq:guess_for_partition_function}
  \begin{aligned}
    m_{ij} &= \frac{(p+q)i + (p-q)j}{p} + 1, \\
    I_p &= \{ (i,j) \in \bZ^2_{\geq 0} \vert i-j = 0 \mod p\}, \\
    P(t; Y^{p,q}) &= \sum_{I_p} m_{ij}
    t^{\frac{\left[3(p+q)-\ell^{-1}\right]i + \left[3(p-q)+\ell^{-1}\right]j}{2p}}.
  \end{aligned}
\end{equation}
Of course, it would be interesting to verify this starting from
\cite{Benvenuti:2006qr}.

\section{Conclusions}
\label{sec:conclusions}

In this paper, we have studied the perturbative partition function of super
Yang-Mills theories on five-dimensional Sasaki-Einstein manifolds $Y$
following the work of Qiu, Zabzine, and collaborators. Using the
intrinsic structure of $Y$, we argued that the contribution from the
vector multiplet can be calculated in terms of Kohn-Rossi cohomology groups
\eqref{eq:one-loop-det_as_Kohn-Rossis}. Thus, the calculation can be
reduced to a counting problem on the Calabi-Yau cone $C(Y)$ which is
very well understood in the context of AdS/CFT duality. This gives an
alternative approach to that via index theorems previously used in the
literature.

Of course, the disagreement of our results for $Y^{2,0}$ and $Y^{4,0}$
with \cite{Qiu:2013pta} is puzzling; yet this has to be taken in light of
the question whether it is possible to define the theory on an
orbifold in the first place. As we argue in appendix
\ref{sec:super-determinant_of_Qiu_Zabzine}, one can see quickly that 
the result of \cite{Qiu:2013pta} for the super determinant (denoted
there as $P_{\sVec}$) is independent of $p$ for $Y^{p,0}$, which holds
not in our case. However, if one restricts to the case $p > q > 0$,
our examples in sections \ref{sec:examples_y7-3} and
\ref{sec:examples_y2-1} suggest full agreement with
\cite{Qiu:2013pta}. Indeed, when performing the necessary calculations
for various examples, the relevant steps take on a somewhat mechanical
nature that simply needs adapting some parameters. This goes hand in
hand with our guess for the generating function in section
\ref{sec:conjecture_on_generating_functions}.
Assuming that the theory might be well-defined
as it is, it is interesting to note that our results for $Y^{2,0}$ and
$Y^{1,0}$ take an identical form, with half the modes contributing to
the latter having been modded out.

Independently of this, note that for regular Sasaki-Einstein manifolds
$Y$, the tangential Cauchy-Riemann operator can be thought of as an
ordinary Dolbeault operator twisted by a suitable line bundle over the
K\"ahler-Einstein base. I.e.~$\tgdb$ is now related to some
$\bar{\partial}_V$ when acting on forms of fixed charge. The latter
was used in \cite{Kallen:2012cs} to evaluate the super determinant
with an index theorem. Considering this comparision in the context of
generic Sasaki-Einstein manifolds, one sees that it should be possible to
calculate the perturbative partition function in terms of the
equivariant index $\ind_{\pounds_\xi} (\tgdb)$. This can also be
seen by considering \eqref{eq:main_result_using_20-_and_02-forms}.

There are some immediate directions of possible future research, such
as the inclusion of hypermultiplets or the calculation of additional
examples such as del Pezzo surfaces. 
Furthermore, as highlighted earlier, we only considered the Reeb vector
that admits a Sasaki-Einstein metric while the results of
\cite{Qiu:2013pta,Qiu:2013aga} hold for generic choices of
$\xi$. Assuming the validity of
our construction in this general case, one might achieve this
generalization by choosing a different diagonal $U(1)$ in the
generating functions $P(t; Y)$. The choice of Reeb and equivariant
parameters features strongly in the latter of the above references,
where the authors used factorization to conjecture the full,
non-perturbative form of the partition function. In general, any
use of the methods employed here towards a better understanding of
contact instantons and the full, non-perturbative partifion function
is of obvious great interest.

\section*{Acknowledgements}

I would like to thank Diego Rodr\'iguez-G\'omez for the many
discussions without which the completion of this project would have
taken considerably longer. Furthermore, I would like to thank Andr\'es
Vi\~na Escalar, Eoin \'O Colg\'ain, Yolanda Lozano, and Maxim Zabzine for various
discussions, comments on the manuscript, and very helpful correspondence respectively.

\appendix

\section{Hodge duals on Sasaki-Einstein manifolds}
\label{sec:technical-aspects}

We review some notation from \cite{Schmude:2013dua} that is quite useful
when manipulating expressions involving the Hodge star operator. The
material is a straightforward generalization of identical ideas on
K\"ahler manifolds to the Sasaki-Einstein case. To begin, we define
the adjoint of the Lefschetz operator $L \equiv  J \wedge$ as well as an
adjoint for the action of the Reeb $L_\eta \equiv \eta \wedge$:
\begin{equation}
  \Lambda = L^* = J \lrcorner, \qquad
  \Lambda_\eta = L_\eta^* = \imath_\xi.
\end{equation}
The space of horizontal forms can be denoted as $\bigoplus
\Omega^{p,q} = \bigwedge^* D^*$. For elements of this space, we
introduce the operator
\begin{equation}
  \boldsymbol{I} = \sum_{p,q} \imath^{p-q} \Pi^{p,q},  
\end{equation}
which uses the projection $\Pi^{p,q}:\Omega^*_{\bC} \to
\Omega^{p,q}$. Finally, we can introduce a restricted Hodge dual
$\bullet$ that acts only on $\bigwedge^* D^*$. The first useful
relation we find is
\begin{equation}\label{eq:se_hodge_identities}
    \star \vert_{\bigwedge^* D^*} = L_\eta \bullet, \qquad
    \star \vert_{\bigwedge^* D^* \wedge \eta} = \bullet (-1)^{d^0} \Lambda_\eta.
\end{equation}
Where
\begin{equation}
  d^0\vert_{\bigwedge^k D^* \wedge (1 \oplus \eta)} = k \cdot \id  
\end{equation}
yields the horizontal degree of a form. We also introduce $P^k = \{
\alpha \in \bigwedge^k D^* \vert \Lambda \alpha = 0\}$, the set of
primitive $k$-forms. With all this notation, one can introduce
Lefschetz decomposition. Given any $\alpha \in \bigwedge^k D^*$, there
is a unique decomposition
\begin{equation}
  \alpha = \sum_r L^r \alpha_r, \qquad
  \alpha_r \in P^{k-2r}.
\end{equation}
Moreover, one can prove the identity
\begin{equation}\label{eq:horizontal_hodge_identity}
  \forall \alpha \in P^k, \qquad
  \bullet L^j \alpha = (-1)^{\frac{k(k-1)}{2}} \frac{j!}{(n-k-j)!} L^{n-k-j} \boldsymbol{I}(\alpha),
\end{equation}
where $d = 2n+1$ is the dimension of the Sasaki-Einstein
manifold. Together with \eqref{eq:se_hodge_identities} this allows for
an efficient evaluation of Hodge duals. The complete algebra involving
$\tgd, \tgdb, L, L_\eta, \pounds_\xi$ and their adjoints was derived in \cite{Schmude:2013dua}.

\section{The super determinant as computed by Qiu and Zabzine }
\label{sec:super-determinant_of_Qiu_Zabzine}

We summarize the result for the one-loop contribution to the partition
function on $Y^{p,q}$ from \cite{Qiu:2013pta}. Again, we restrict to
the abelian case
\begin{equation}
  \sdet^\prime(-\imath L_\xi) = \prod_{\Lambda_0^+} (i \omega_1 + j \omega_2 + k
  \omega_3 + l \omega_4)^2 \prod_{\Lambda_1^+} (i\omega_1 + j \omega_2
  + k \omega_3 + l \omega_4).
\end{equation}
The integers $i, j, k, l$ lie in the lattices
\begin{equation}
  \begin{aligned}
    \Lambda^+ &= \{i, j, k, l \in \bZ_{\geq 0} \vert (p+q)i +(p-q) j =
    p(k+l) \}, \\
    \Lambda^+_0 &= \{i, j, k, l \in \bZ_{> 0} \vert (p+q)i +(p-q) j =
    p(k+l) \}, \\
    \Lambda^+_1 &= \Lambda^+ \setminus (\Lambda^+_0 \cup \{0, 0, 0, 0\}).
  \end{aligned}
\end{equation}
The $\omega_i$ depend on the choice of Reeb with the supersymmetric
choice being
\begin{equation}
  \begin{aligned}
    \omega_1 = 0, \qquad
    \omega_2 = \frac{\ell^{-1}}{p+q}, \qquad
    \omega_3 = \omega_4 = \frac{3}{2} - \frac{\ell^{-1}}{2(p+q)}.
  \end{aligned}
\end{equation}
$\ell^{-1}$ was defined in equation \eqref{eq:ell_inverse}.

For $Y^{1,0} = T^{1,1}$, we define $n \equiv k+l$ and note that the number of
lattice points for fixed $n$ is
\begin{equation}\label{eq:t11_lattices}
  \#\Lambda^+\vert_{n \geq 0} = (n+1)^2, \qquad
  \#\Lambda^+_0\vert_{n>0} = (n-1)^2, \qquad
  \#\Lambda^+_1\vert_{n\geq 0} = 4n.
\end{equation}
Upon substitution, this confirms \eqref{eq:t11_result}. As a matter of
fact, the lattices \eqref{eq:t11_lattices} are identical for all
$Y^{p,0}$ since $p$ simply drops out. The same holds for the $\omega_i$.

For $Y^{7,3}$, we note that the integers $i, j, k, l$ are subject to
the constraint
\begin{equation}
  10 i + 4j = 7(k+l).
\end{equation}
We introduce the set
\begin{equation}
  I_7 = \{(i, j) \in \bZ^2_{\geq 0} \vert i-j = 0 \mod 7\}.  
\end{equation}
For a pair $(i, j) \in I_7$, we find that $\Lambda_0^+\vert_{(i,j)}$
contains $\frac{10i+4j}{7} - 1$ and
$\Lambda^+\vert_{(i,j)}$ $\frac{10i+4j}{7}+1$ lattice points. If $i =
0$ or $j = 0$, $\Lambda_1^+\vert_{(i,j)}$ consists also of
$\frac{10i+4j}{7}+1$ points, yet if $i j \neq 0$, there are only two
points in $\Lambda_1^+\vert_{(i,j)}$. To calculate the
super-determinant, we eliminate $k+l$ and find
\begin{equation}
  \sdet^\prime(-\imath \pounds_\xi) = \prod_{I_7 \vert i, j > 0}
  \left( \frac{5i+4j}{3} \right)^{2 \frac{10i+4j}{7}}
  \cdot
  \prod_{I_7 \vert i = 0 \lor j = 0} \left(\frac{5i+4j}{3}\right)^{\frac{10i+4j}{7}+1}.
\end{equation}

For $Y^{2,1}$ the situation is almost identical. Here we define
\begin{equation}
  I_2 = \{(i,j) \in \bZ^2_{\geq 0} \vert i-j=0\mod 2\}
\end{equation}
to paremetrize the lattices. Things work out in a way identical to
$Y^{7,3}$ and one finds
\begin{IEEEeqnarray}{rCl}
  \sdet(-\imath\pounds_\xi) = & \prod_{I_2 \vert i, j > 0} & \left(
    \frac{(7-\sqrt{13})i + (\sqrt{13}-1)j}{2} \right)^{3i+j} \nonumber\\
  &\prod_{I_2 \vert i = 0 \lor j = 0}& \left( \frac{(7-\sqrt{13})i + (\sqrt{13}-1)j}{2}
  \right)^{\frac{3i+j}{2} +1}. \label{eq:y21_result}
\end{IEEEeqnarray}

\section{$S^5$ and the Borel-Weil-Bott theorem}
\label{sec:s5-borel-weil-bott}

Since $S^5$ is a regular Sasaki-Einstein manifold, the orbits of the
Reeb close and yield a principal bundle over $\CP^2$. It follows that
the Kohn-Rossi cohomology groups with fixed charge $n$ are isomorphic
to the cohomology groups of the base twisted by a suitable line bundle.
Then, the Borel-Weil-Bott theorem\footnote{See the appendix of
  \cite{Eager:2013mua} for further examples of this.}
allows us to relate these to representations of $A_2$.
\begin{equation}
  \begin{aligned}
    H_\tgdb^{0,0}(S^5) \vert_n &\cong H^0(\CP^2, \lineB^n) \cong
    V^{A_2}_{\lbrack n, 0, 0\rbrack}, \\
    H_\tgdb^{2,0}(S^5) \vert_n &\cong H^0(\CP^2, \Omega^2 \otimes
    \lineB^n) \cong V^{A_2}_{\lbrack n-3, 0, 0\rbrack}.
  \end{aligned}
\end{equation}
Finally, we use the Weyl dimension formula\footnote{See e.g.~equation 7.18 in
  \cite{GoodmanWallach2009}.}
\begin{equation}
  \dim V_\lambda = \prod_{1 \leq i < j \leq n} \frac{\lambda_i -
    \lambda_j + j - i}{j-i}
\end{equation}
to calculate the dimension of the cohomology groups:
\begin{equation}
  \begin{aligned}
    \dim V_{\lbrack n, 0, 0\rbrack} &= 1 + \frac{3}{2} n + \frac{1}{2}
    n^2, &\qquad &(=\ind \bar{\partial}_V), \\
    \dim V_{\lbrack n-3, 0, 0\rbrack} &= 1 - \frac{3}{2} n + \frac{1}{2}
    n^2, &\qquad &(=\ind \partial_V).
  \end{aligned}
\end{equation}
The indices $\ind \bar{\partial}_V$ and $\ind \partial_V$ were calculated in \cite{Kallen:2012cs}.

\bibliographystyle{ytphys}
\small\baselineskip=.97\baselineskip
\bibliography{ref}

\providecommand{\href}[2]{#2}\begingroup\raggedright\begin{thebibliography}{10}

\bibitem{Witten:1988ze}
E.~Witten, ``{Topological Quantum Field Theory},''
\href{http://dx.doi.org/10.1007/BF01223371}{{\em Commun.Math.Phys.} {\bfseries
  117} (1988) 353}.

\bibitem{Witten:1991zz}
E.~Witten, ``{Mirror manifolds and topological field theory},''
\href{http://arxiv.org/abs/hep-th/9112056}{{\ttfamily arXiv:hep-th/9112056
  [hep-th]}}.

\bibitem{Beasley:2005vf}
C.~Beasley and E.~Witten, ``{Non-Abelian localization for Chern-Simons
  theory},'' {\em J.Diff.Geom.} {\bfseries 70} (2005) 183--323,
\href{http://arxiv.org/abs/hep-th/0503126}{{\ttfamily arXiv:hep-th/0503126
  [hep-th]}}.

\bibitem{Pestun:2007rz}
V.~Pestun, ``{Localization of gauge theory on a four-sphere and supersymmetric
  Wilson loops},'' \href{http://dx.doi.org/10.1007/s00220-012-1485-0}{{\em
  Commun.Math.Phys.} {\bfseries 313} (2012) 71--129},
\href{http://arxiv.org/abs/0712.2824}{{\ttfamily arXiv:0712.2824 [hep-th]}}.

\bibitem{Doroud:2012xw}
N.~Doroud, J.~Gomis, B.~Le~Floch, and S.~Lee, ``{Exact Results in D=2
  Supersymmetric Gauge Theories},''
  \href{http://dx.doi.org/10.1007/JHEP05(2013)093}{{\em JHEP} {\bfseries 1305}
  (2013) 093},
\href{http://arxiv.org/abs/1206.2606}{{\ttfamily arXiv:1206.2606 [hep-th]}}.

\bibitem{Benini:2013xpa}
F.~Benini, R.~Eager, K.~Hori, and Y.~Tachikawa, ``{Elliptic genera of 2d N=2
  gauge theories},''
\href{http://arxiv.org/abs/1308.4896}{{\ttfamily arXiv:1308.4896 [hep-th]}}.

\bibitem{Kim:2013ola}
H.~Kim, S.~Lee, and P.~Yi, ``{Exact Partition Functions on RP2 and
  Orientifolds},''
\href{http://arxiv.org/abs/1310.4505}{{\ttfamily arXiv:1310.4505 [hep-th]}}.

\bibitem{Kapustin:2009kz}
A.~Kapustin, B.~Willett, and I.~Yaakov, ``{Exact Results for Wilson Loops in
  Superconformal Chern-Simons Theories with Matter},''
  \href{http://dx.doi.org/10.1007/JHEP03(2010)089}{{\em JHEP} {\bfseries 1003}
  (2010) 089},
\href{http://arxiv.org/abs/0909.4559}{{\ttfamily arXiv:0909.4559 [hep-th]}}.

\bibitem{Imamura:2011wg}
Y.~Imamura and D.~Yokoyama, ``{N=2 supersymmetric theories on squashed
  three-sphere},'' \href{http://dx.doi.org/10.1103/PhysRevD.85.025015}{{\em
  Phys.Rev.} {\bfseries D85} (2012) 025015},
\href{http://arxiv.org/abs/1109.4734}{{\ttfamily arXiv:1109.4734 [hep-th]}}.

\bibitem{Hama:2011ea}
N.~Hama, K.~Hosomichi, and S.~Lee, ``{SUSY Gauge Theories on Squashed
  Three-Spheres},'' \href{http://dx.doi.org/10.1007/JHEP05(2011)014}{{\em JHEP}
  {\bfseries 1105} (2011) 014},
\href{http://arxiv.org/abs/1102.4716}{{\ttfamily arXiv:1102.4716 [hep-th]}}.

\bibitem{Nian:2013qwa}
J.~Nian, ``{Localization of Supersymmetric Chern-Simons-Matter Theory on a
  Squashed $S^3$ with $SU(2)\times U(1)$ Isometry},''
\href{http://arxiv.org/abs/1309.3266}{{\ttfamily arXiv:1309.3266 [hep-th]}}.

\bibitem{Tanaka:2013dca}
A.~Tanaka, ``{Localization on round sphere revisited},''
  \href{http://dx.doi.org/10.1007/JHEP11(2013)103}{{\em JHEP} {\bfseries 1311}
  (2013) 103},
\href{http://arxiv.org/abs/1309.4992}{{\ttfamily arXiv:1309.4992 [hep-th]}}.

\bibitem{Fujitsuka:2013fga}
M.~Fujitsuka, M.~Honda, and Y.~Yoshida, ``{Higgs branch localization of 3d N=2
  theories},''
\href{http://arxiv.org/abs/1312.3627}{{\ttfamily arXiv:1312.3627 [hep-th]}}.

\bibitem{Alday:2013lba}
L.~F. Alday, D.~Martelli, P.~Richmond, and J.~Sparks, ``{Localization on
  Three-Manifolds},''
\href{http://arxiv.org/abs/1307.6848}{{\ttfamily arXiv:1307.6848 [hep-th]}}.

\bibitem{Gomis:2011pf}
J.~Gomis, T.~Okuda, and V.~Pestun, ``{Exact Results for 't Hooft Loops in Gauge
  Theories on $S^4$},'' \href{http://dx.doi.org/10.1007/JHEP05(2012)141}{{\em
  JHEP} {\bfseries 1205} (2012) 141},
\href{http://arxiv.org/abs/1105.2568}{{\ttfamily arXiv:1105.2568 [hep-th]}}.

\bibitem{Closset:2013sxa}
C.~Closset and I.~Shamir, ``{The $\mathcal{N}=1$ Chiral Multiplet on $T^2\times
  S^2$ and Supersymmetric Localization},''
\href{http://arxiv.org/abs/1311.2430}{{\ttfamily arXiv:1311.2430 [hep-th]}}.

\bibitem{Kim:2012ava}
H.-C. Kim and S.~Kim, ``{M5-branes from gauge theories on the 5-sphere},''
  \href{http://dx.doi.org/10.1007/JHEP05(2013)144}{{\em JHEP} {\bfseries 1305}
  (2013) 144},
\href{http://arxiv.org/abs/1206.6339}{{\ttfamily arXiv:1206.6339 [hep-th]}}.

\bibitem{Imamura:2012bm}
Y.~Imamura, ``{Perturbative partition function for squashed $S^5$},''
\href{http://arxiv.org/abs/1210.6308}{{\ttfamily arXiv:1210.6308 [hep-th]}}.

\bibitem{Kim:2012qf}
H.-C. Kim, J.~Kim, and S.~Kim, ``{Instantons on the 5-sphere and M5-branes},''
\href{http://arxiv.org/abs/1211.0144}{{\ttfamily arXiv:1211.0144 [hep-th]}}.

\bibitem{Kallen:2012cs}
J.~K{\"a}ll{\'e}n and M.~Zabzine, ``{Twisted supersymmetric 5D Yang-Mills
  theory and contact geometry},''
  \href{http://dx.doi.org/10.1007/JHEP05(2012)125}{{\em JHEP} {\bfseries 1205}
  (2012) 125},
\href{http://arxiv.org/abs/1202.1956}{{\ttfamily arXiv:1202.1956 [hep-th]}}.

\bibitem{Kallen:2012va}
J.~K{\"a}ll{\'e}n, J.~Qiu, and M.~Zabzine, ``{The perturbative partition
  function of supersymmetric 5D Yang-Mills theory with matter on the
  five-sphere},'' \href{http://dx.doi.org/10.1007/JHEP08(2012)157}{{\em JHEP}
  {\bfseries 1208} (2012) 157},
\href{http://arxiv.org/abs/1206.6008}{{\ttfamily arXiv:1206.6008 [hep-th]}}.

\bibitem{Qiu:2013pta}
J.~Qiu and M.~Zabzine, ``{5D Super Yang-Mills on $Y^{p,q}$ Sasaki-Einstein
  manifolds},''
\href{http://arxiv.org/abs/1307.3149}{{\ttfamily arXiv:1307.3149 [hep-th]}}.

\bibitem{Qiu:2013aga}
J.~Qiu and M.~Zabzine, ``{Factorization of 5D super Yang-Mills on $Y^{p,q}$
  spaces},''
\href{http://arxiv.org/abs/1312.3475}{{\ttfamily arXiv:1312.3475 [hep-th]}}.

\bibitem{Hama:2010av}
N.~Hama, K.~Hosomichi, and S.~Lee, ``{Notes on SUSY Gauge Theories on
  Three-Sphere},'' \href{http://dx.doi.org/10.1007/JHEP03(2011)127}{{\em JHEP}
  {\bfseries 1103} (2011) 127},
\href{http://arxiv.org/abs/1012.3512}{{\ttfamily arXiv:1012.3512 [hep-th]}}.

\bibitem{Hosomichi:2012ek}
K.~Hosomichi, R.-K. Seong, and S.~Terashima, ``{Supersymmetric Gauge Theories
  on the Five-Sphere},''
  \href{http://dx.doi.org/10.1016/j.nuclphysb.2012.08.007}{{\em Nucl.Phys.}
  {\bfseries B865} (2012) 376--396},
\href{http://arxiv.org/abs/1203.0371}{{\ttfamily arXiv:1203.0371 [hep-th]}}.

\bibitem{Festuccia:2011ws}
G.~Festuccia and N.~Seiberg, ``{Rigid Supersymmetric Theories in Curved
  Superspace},'' \href{http://dx.doi.org/10.1007/JHEP06(2011)114}{{\em JHEP}
  {\bfseries 1106} (2011) 114},
\href{http://arxiv.org/abs/1105.0689}{{\ttfamily arXiv:1105.0689 [hep-th]}}.

\bibitem{Closset:2012ru}
C.~Closset, T.~T. Dumitrescu, G.~Festuccia, and Z.~Komargodski,
  ``{Supersymmetric Field Theories on Three-Manifolds},''
  \href{http://dx.doi.org/10.1007/JHEP05(2013)017}{{\em JHEP} {\bfseries 1305}
  (2013) 017},
\href{http://arxiv.org/abs/1212.3388}{{\ttfamily arXiv:1212.3388 [hep-th]}}.

\bibitem{Closset:2013vra}
C.~Closset, T.~T. Dumitrescu, G.~Festuccia, and Z.~Komargodski, ``{The Geometry
  of Supersymmetric Partition Functions},''
\href{http://arxiv.org/abs/1309.5876}{{\ttfamily arXiv:1309.5876 [hep-th]}}.

\bibitem{Nieri:2013yra}
F.~Nieri, S.~Pasquetti, and F.~Passerini, ``{3d \& 5d gauge theory partition
  functions as q-deformed CFT correlators},''
\href{http://arxiv.org/abs/1303.2626}{{\ttfamily arXiv:1303.2626 [hep-th]}}.

\bibitem{Nieri:2013vba}
F.~Nieri, S.~Pasquetti, F.~Passerini, and A.~Torrielli, ``{5D partition
  functions, q-Virasoro systems and integrable spin-chains},''
\href{http://arxiv.org/abs/1312.1294}{{\ttfamily arXiv:1312.1294 [hep-th]}}.

\bibitem{Lockhart:2012vp}
G.~Lockhart and C.~Vafa, ``{Superconformal Partition Functions and
  Non-perturbative Topological Strings},''
\href{http://arxiv.org/abs/1210.5909}{{\ttfamily arXiv:1210.5909 [hep-th]}}.

\bibitem{Eager:2012hx}
R.~Eager, J.~Schmude, and Y.~Tachikawa, ``{Superconformal Indices,
  Sasaki-Einstein Manifolds, and Cyclic Homologies},''
\href{http://arxiv.org/abs/1207.0573}{{\ttfamily arXiv:1207.0573 [hep-th]}}.

\bibitem{Eager:2013mua}
R.~Eager and J.~Schmude, ``{Superconformal Indices and M2-Branes},''
\href{http://arxiv.org/abs/1305.3547}{{\ttfamily arXiv:1305.3547 [hep-th]}}.

\bibitem{Schmude:2013dua}
J.~Schmude, ``{Laplace operators on Sasaki-Einstein manifolds},''
\href{http://arxiv.org/abs/1308.1027}{{\ttfamily arXiv:1308.1027 [hep-th]}}.

\bibitem{Benvenuti:2006qr}
S.~Benvenuti, B.~Feng, A.~Hanany, and Y.-H. He, ``{Counting BPS Operators in
  Gauge Theories: Quivers, Syzygies and Plethystics},''
  \href{http://dx.doi.org/10.1088/1126-6708/2007/11/050}{{\em JHEP} {\bfseries
  0711} (2007) 050},
\href{http://arxiv.org/abs/hep-th/0608050}{{\ttfamily arXiv:hep-th/0608050
  [hep-th]}}.

\bibitem{Sparks:2010sn}
J.~Sparks, ``{Sasaki-Einstein Manifolds},''
  \href{http://dx.doi.org/10.4310/SDG.2011.v16.n1.a6}{{\em Surveys Diff.Geom.}
  {\bfseries 16} (2011) 265--324},
\href{http://arxiv.org/abs/1004.2461}{{\ttfamily arXiv:1004.2461 [math.DG]}}.

\bibitem{Boyer:2004fc}
C.~P. Boyer and K.~Galicki, ``{Sasakian geometry, hypersurface singularities,
  and Einstein metrics},''
\href{http://arxiv.org/abs/math/0405256}{{\ttfamily arXiv:math/0405256
  [math-dg]}}.

\bibitem{Gauntlett:2006vf}
J.~P. Gauntlett, D.~Martelli, J.~Sparks, and S.-T. Yau, ``{Obstructions to the
  existence of Sasaki-Einstein metrics},''
  \href{http://dx.doi.org/10.1007/s00220-007-0213-7}{{\em Commun.Math.Phys.}
  {\bfseries 273} (2007) 803--827},
\href{http://arxiv.org/abs/hep-th/0607080}{{\ttfamily arXiv:hep-th/0607080
  [hep-th]}}.

\bibitem{Kallen:2011ny}
J.~Kallen, ``{Cohomological localization of Chern-Simons theory},''
  \href{http://dx.doi.org/10.1007/JHEP08(2011)008}{{\em JHEP} {\bfseries 1108}
  (2011) 008},
\href{http://arxiv.org/abs/1104.5353}{{\ttfamily arXiv:1104.5353 [hep-th]}}.

\bibitem{Martelli:2004wu}
D.~Martelli and J.~Sparks, ``{Toric geometry, Sasaki-Einstein manifolds and a
  new infinite class of AdS/CFT duals},''
  \href{http://dx.doi.org/10.1007/s00220-005-1425-3}{{\em Commun.Math.Phys.}
  {\bfseries 262} (2006) 51--89},
\href{http://arxiv.org/abs/hep-th/0411238}{{\ttfamily arXiv:hep-th/0411238
  [hep-th]}}.

\bibitem{Martelli:2005tp}
D.~Martelli, J.~Sparks, and S.-T. Yau, ``{The Geometric dual of a-maximisation
  for Toric Sasaki-Einstein manifolds},''
  \href{http://dx.doi.org/10.1007/s00220-006-0087-0}{{\em Commun.Math.Phys.}
  {\bfseries 268} (2006) 39--65},
\href{http://arxiv.org/abs/hep-th/0503183}{{\ttfamily arXiv:hep-th/0503183
  [hep-th]}}.

\bibitem{Bertolini:2004xf}
M.~Bertolini, F.~Bigazzi, and A.~Cotrone, ``{New checks and subtleties for
  AdS/CFT and a-maximization},''
  \href{http://dx.doi.org/10.1088/1126-6708/2004/12/024}{{\em JHEP} {\bfseries
  0412} (2004) 024},
\href{http://arxiv.org/abs/hep-th/0411249}{{\ttfamily arXiv:hep-th/0411249
  [hep-th]}}.

\bibitem{GoodmanWallach2009}
R.~Goodman and N.~R. Wallach, {\em Symmetry, Representations, and Invariants}.
\newblock Graduate Texts in Mathematics. Springer, New York, 2009.

\end{thebibliography}\endgroup

\end{document}